# Study on the Fermi level of microstructured Silicon with impurities introduced by chalcogenides and their affect on solar cell efficiency

Huili He, Changshui Chen*, Fang Wang, Songhao Liu

*Abstract*—Microstructured Silicon, which is obtained by irradiating the surface of a Silicon wafer with femtosecond laser pulses under certain circumstances, has unusual optical properties such as the strong absorption of light with wavelength from *0.25μm* to *17μm*. So it holds great promise in the intermediate band solar cell (IBSC). Some articles have discussed the electronic structure associating with simple substitutional impurities in Silicon introduced by chalcogenides. And on this basis, after high temperature annealing treatment, we establish the mode of impurity levels of microstructured Silicon introduced by sulfur and oxygen. Using generalized statistics of multi-level, we analyze the probability of electronic in all local energy levels and the relationship among Fermi level, temperature and the density of impurities. Then the theoretical conversion efficiency of the corresponding IBSC is discussed with the Detailed Balance Theory. And the issue of making high efficiency solar cells based on femtosecond laser microstructured Silicon is discussed in detail.

## I. Introduction

The energy crisis and the requirement for environmental protection have made the research of high-efficiency solar cell to be the focus of global technical workers' attention. The research of all kinds of new solar cells is to improve photoelectric conversion efficiency and reduce the production cost; and the two major objectives are exploring actively in the developed countries and some developing countries. The solar cell which is made with new materials, new structures and new craft manufacture is called new solar cell. Now the two focus problems and priorities of the research and development on global new solar cell are crystalline silicon high efficient solar cells and all kinds of thin film solar cells.

In 1997, repeated pulsed laser irradiation of silicon in the presence of sulfur-bearing gases has been used to create microstructured silicon surfaces Professor Eric Mazur in Hazard University named the new material "black silicon"[1]. Researchers found that this new sort of material had extraordinary optical properties to name a few, the strong absorption of light within the range of 0.25μm to 17μm [2-7], nice field emission characteristics, etc, Which provide silicon with various new features. Professor Mazur had predicted that microstructured silicon had incomparable superiority to other materials for use in the solar cell field. Besides, these materials have important potential applications in the fields of detector, solar sell, sensor, display technology, microelectronics and so on. The two most striking features of microstructured Silicon are its extremely low reflection of sunlight and its broad-spectrum absorption (about 17μm wide). These are just its unique advantages to make detector, solar cell and so on. Research results have shown that the enhancement of visible light absorption of microstructured silicon is mainly attributed to tiny "islands and pools" named by the periodic structure on its surface. And compared with the same amount of other materials, the microstructured silicon can absorb more than 42% of the light. Thus, nanoscale texture makes it possible to manufacture efficient solar cells with few materials [8]; while the strong infrared absorption is mainly due to the deep levels and dark spot defects in the band gap formed in the ultra-fast laser machining process [9-11]. Chalcogen O, S, Se, and Te are dramatic impurities in Silicon [12, 13], and they have potential to increase the transformation efficiency of photovoltaic solar cell. It is widely believed that chalcogenides belong to substitutional impurities, although there are some controversies in this filed. Recently, domestic and international researchers have reported that Chalcogenides in silicon can produce impurities levels in silicon band gap, because of the gradual improvement of the material manufacturing technology and the development of the various spectrum technology including electronics transient spectrum, light absorption, light conductance, magnetic resonance and so on.

Fig.1 shows the reported impurity levels in Silicon introduced by sulfur and oxygen[13]. And these impurity levels

Manuscript received February 6, 2012. This work was co-funded by the Key Program of Natural Science Foundation of Guangdong province (No. 10251063101000001; No.8251063101000006) and the National Natural Science Foundation of China (No. 60878063)

The authors are with School of Information and Optoelectronic Science and Engineering，South China Normal University，Guangzhou 510631, Guangdong, China. (e-mail: hehuili05@163.com; e-mail: cschen@aiofm.ac.cn; e-mail: tiantangniao1205@163.com; e-mail: liush@scnu.edu.cn).
corresponding author: Fax:0086+20+85211768, phone: 0086+20+85211768, e-mail: cschen@aiofm.ac.cn;

have not undergone annealing treatment. Energy counts from the bottom of conduction band (BCB), and its unit is meV.

Oxygen is found abundant in Silicon. Amount of Oxygen is determined by crystal growth, and its existing form and content changes with annealing temperature and time. The following is the existing form of sulfur changing with the increasing temperature:

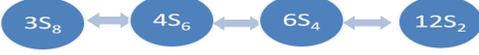

And on this basis, after high temperature annealing, the model of impurity level of microstructured Silicon introduced by sulfur and oxygen is established as shown in Fig. 2. $E_{D3}$ and $E_A$ are the donor level and acceptor level introduced by oxygen, and the other impurity levels are introduced by sulfur.

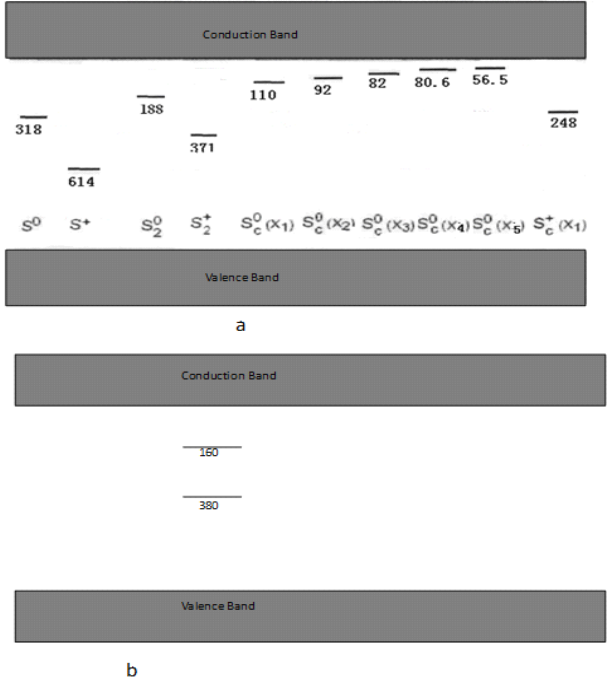

FIG.1. Impurity levels in Silicon introduced by sulfur and oxygen,(a)S,(b)O.

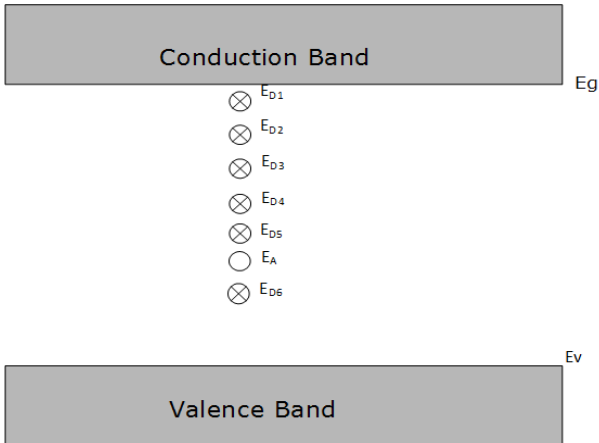

FIG.2. Impurity level mode of microstructured Silicon introduced by sulfur and oxygen. $E_{D1}$=1.028eV, $E_{D2}$=1.01eV, $E_{D3}$=0.96eV, $E_{D4}$=0.932eV, $E_{D5}$=0.802eV, $E_{D6}$=0.506eV, $E_A$=0.74eV.

In conformity to the sequence of the impurity levels in Fig.2, we analyze the probability of electronic in all local energy levels and the relationship among Fermi level, temperature and the density of impurities in microstructured Silicon. Then we discuss the conversion efficiency of the IBSC based on this material.

## II. THE RELATIONSHIP AMONG FERMI LEVEL, TEMPERATURE AND THE DENSITY OF IMPURITIES IN MICROSTRUCTURED SILICON

The distribution rate of each local level is not unrelated obviously. Each level would be occupied by a certain number of impurities from the total, and the sum of all these numbers plus the number of unionized impurity centers should keep 1 [13]. The occupation rate of a given level can not be directly obtained from the Fermi function, and the statistical relationship of thermal equilibrium occupancy of two neighboring states is required [13, 14].

$$\frac{N_s}{N_{s-1}} = \frac{g_{s-1}}{g_s} \exp(\frac{E_F - E_s}{kT}) \quad (1)$$

The meaning of the letters in the formula can be seen from reference 15. The letter g is the corresponding degeneracy of spin energy, with its value of 1 or 2. The corresponding value of g can be determined by the state changes caused by an electronic removing from a state to one of two possible states. So, if S is an acceptor level, then

$$N_s = \frac{1}{2} N_{s-1} \exp(\frac{E_F - E_s}{kT}) \quad (2)$$

Similarly, if $S$-1 is a donor level, then

$$N_{s-1} = \frac{1}{2} N_s \exp(\frac{E_{s-1} - E_F}{kT}) \quad (3)$$

Calculated according to equilibrium statistical, the density of impurity center in semiconductor materials is:

$$N = N_0 + N_A + \sum_i^5 N_{Di} \quad (4)$$

Where $N_A$ is the density of impurity center in corresponding level $E_A$; $N_{Di}$ is the density of impurity center in corresponding level $E_{Di}$, According to Eq. (2), we can get:

$$N_A = \frac{1}{2} N_0 \exp(\frac{E_F - E_A}{kT}) \quad (5)$$

According to Eq. (3), we can get:

$$\begin{cases} N_{D1} = \frac{1}{2} N_0 \exp(\frac{E_{D1} - E_F}{kT}) \\ N_{D2} = \frac{1}{2} N_{D1} \exp(\frac{E_{D2} - E_F}{kT}) = \left(\frac{1}{2}\right)^2 N_0 \exp(\frac{E_{D1} - E_F}{kT}) \exp(\frac{E_{D2} - E_F}{kT}) \\ \dots \\ N_{D6} = \frac{1}{2} N_{D5} \exp(\frac{E_{D6} - E_F}{kT}) = \left(\frac{1}{2}\right)^6 N_0 \prod_{i=1}^{6} \exp(\frac{E_{Di} - E_F}{kT}) \end{cases} \quad (6)$$

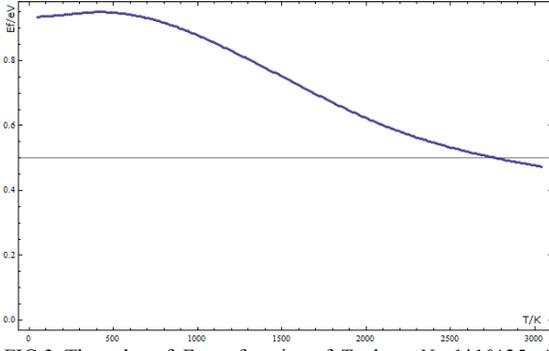

FIG.3. The value of $E_f$ as a function of $T$, where $N_o=1*10^{25}$m$^{-3}$ and $N_s=7*10^{25}$m$^{-3}$.

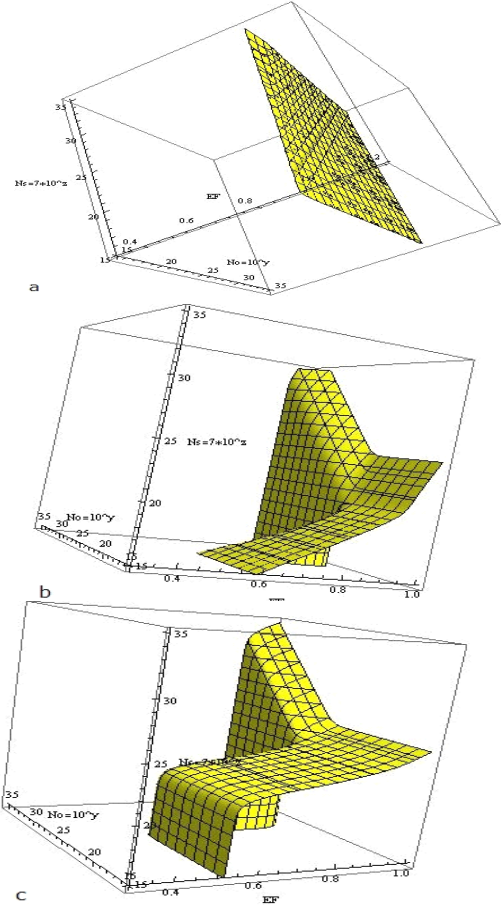

FIG.4. The relationship among $E_F$, $N_o$ and $N_s$, where T=78K(a), T=300K(b) and T=1000(c)

With Eqs. (5) and (6), Eq. (4) should be changed into a more precise form:

$$N = N_0 + \frac{1}{2}N_0 \exp(\frac{E_F - E_A}{kT}) + N_0 \sum_{l=1}^{6}\left(\frac{1}{2}\right)^l \left[\prod_{i=1}^{l} \exp(\frac{E_{Di} - E_F}{kT})\right] \quad (7)$$

Now, we can define a coefficient "$D$" as follow:

$$N = N_0 + \frac{1}{2}N_0 \exp(\frac{E_F - E_A}{kT}) + N_0 \sum_{l=1}^{6}\left(\frac{1}{2}\right)^l \left[\prod_{i=1}^{l} \exp(\frac{E_{Di} - E_F}{kT})\right] \quad (8)$$

$$F(E_A) \equiv \frac{N_A}{N} = \frac{\frac{1}{2}\exp(\frac{E_F - E_A}{kT})}{D} \quad (9)$$

$$F(E_{Dq}) \equiv \frac{N'_{Dq}}{N} \quad (10)$$

Where $N'_{Dq} = \sum_{a=q}^{6} N_{Da}$.

Then according to the condition of electrical neutrality, we can gets:

$$p + N_D^+ = n + N_A^- \quad (11)$$

Where $p$ is the hole density in valence band; $n$ is the electronic density in conduction band; $N_A^-$ is the bound electron density in the impurity center ($N_A^- = N_A F(E_A)$); $N_D^+$ is the bound hole density in the impurity center ($N_D^+ = \sum_{q=1}^{2}[N_{Dq}\sum_{i=1}^{m_q}F(E_{Di})]$).

Then according to all the above, Eq. (11) should be changed into a more precise form:

$$p - n = N_A^- - N_D^+ = N_O F(E_A) - [N_O F(E_{DO}) + N_S \sum_{i=1}^{4} F(E_{Dsi})] \quad (12)$$

Where $N_o$ is the density of impurities introduced by oxygen in the microstructured Silicon; $N_s$ is the density of impurities introduced by sulfur in the microstructured Silicon.

To form a stable intermediate impurity band in the forbidden gap of Silicon, there should require the doping concentration of impurities not lower than $10^{25}$ m$^{-3}$. The general doping concentration of sulfur in femtosecond laser microstructured Silicon is $10^{25}$ or $10^{26}$ orders of magnitude so that femtosecond laser microstructured Silicon can meet the requirement for concentration.

Calculating by the formula above, we have the relationship among Fermi level, Temperature and the Density of Impurities in the microstructured Silicon [Fig.3, Fig.4].

Fig.3 shows the relationship between $E_f$ and $T$, when $N_s=7*10^{25}$ m$^{-3}$ and $N_o=1*10^{25}$ m$^{-3}$. The figure demonstrates that Fermi energy increases with temperature rising in low temperature area. However, at the temperature that is beyond 500K, Fermi energy decreases with temperature rising and decreases very slowly with temperature rising in the extremely high temperature. Now we can work out the density of impurity centers in each impurity level at different temperatures by taking the value of Fermi energy at different temperatures into Eqs. (5) and (6).

Fig.4 shows the relationship among the Fermi energy, the doping concentration of sulfur and the doping concentration of oxygen at low temperature, room temperature and high temperature. The figures demonstrate that Fermi energy only has a slight change with the increase of the doping concentration of sulfur and oxygen in a certain range of doping concentration of impurities. Once beyond this doping concentration, Fermi energy has a sharp change with a tiny rise of the doping concentration of impurities, then changes slightly, and has the repeatability. In the room temperature, Fermi energy rises along with the increase of the doping concentration of sulfur and oxygen in the low doping concentration range, and Fermi energy rises slowly in the high doping concentration range (Fig.4.(b)).

According to the analysis above，we can work out the Fermi energy of microstructured Silicon under annealing treatment at different temperatures. Now we can obtain the extent of electronics occupying all the local level by taking the above value of Fermi energy into Eqs. (5) and (6), then the number of free holes and free electrons in the material can be obtained by an appropriate sum method. Besides, we can also calculate the concentration of minority carriers in this situation. So Fermi energy is essential to the physical property of the material such as conductivity, electric capacity and so on, especially for the semiconductor.

### III. ANALYSIS OF THE EFFICIENCY OF THE IBSC BASED ON FEMTOSECOND LASER MICROSTRUCTURED SILICON

The IBSC can, in principle, increase photoelectric conversion efficiency of a conventional single junction solar cell by introducing intermediate bands into the device. They ideally allow electrons to be excited from the valence band to the conduction band via intermediate band by absorbing the previously wasted sub-band-gap photons. In this work, We use Detailed Balance Theory [15-17] to discuss the conversion efficiency of microstructured Silicon solar cell with seven intermediate bands (Fig.5). Fig.5 shows Energy band diagram and circuit diagram of the IPV cell containing seven defect species: (a) energy gaps and generated currents and (b) an equivalent circuit in terms of their associated chemical potential. In Fig.5.(a) the level position starts from the conduction band, with gradual values of 0.092eV, 0.11eV, 0.16eV, 0.188eV, 0.318eV, 0.380eV and 0.614eV. However, a high-energy photon can generate numbers of electron hole pairs for its special structure of femtosecond laser microstructured Silicon. And there is still no comparatively comprehensive understanding of this phenomenon up till now, so it must be noted that we do not consider the phenomenon of multiple electronic outputs for femtosecond laser microstructured Silicon.

When the cell absorbs a photon with sufficient energy, an electron is excited from the valence band to the conduction band. In the device described above, this can occur in two different ways: (1) Only one photon with the energy greater than the band-gap of the cell is needed to excite the electron from the valence band to the conduction band directly. (2) Two photons are needed to excite an electron from the valence band to the conduction band. The first photon excites electron from the valence band to an intermediate band; the second photon is then absorbed to excite the electron from the intermediate band to the conduction band. And four assumptions are made in our calculation as follows:

(a) Impurities in the device are assumed to give rise to energy levels within the forbidden gap of the device; in general, each different impurity species will give rise to a separate energy level. The bandwidth of conduction band and the valence band are not considered.
(b) $0 < E_n < E_{g/2}$ and $E_{n+1} \leq E_n \leq E_{n-1}$ (n=1,2….7).
(c) Photons with energy between $E_g-E_n$ and $E_g-E_{n-1}$ can make the electrons excited from the valence band into the nth energy level while photons with energy between $E_n$ and $E_{n+1}$ can make the electrons excited from the nth intermediate energy level into the conduction band. The current flow from the valence band to the nth intermediate energy level must be equal to the current flow from the nth intermediate energy level to the conduction band ($J_{Vn} = J_{nC}$, n=1,2….7).
(d) Considering purely from the algebraic standpoint, the change in the chemical potential due to transitions from the valence band to conduction band must be equal to the sum of the chemical potential changing from the valence band to the nth intermediate energy level and the chemical potential changing from the nth intermediate energy level to the conduction band ($u_{VC}=u_{Vn}+u_{nC}$, n=1,2….7).

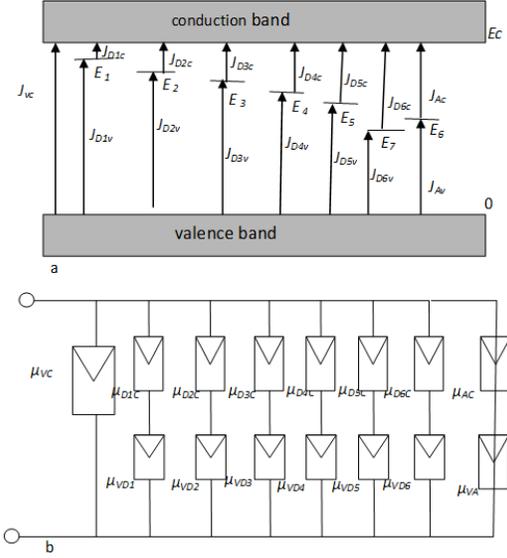

FIG.5. Energy band diagram and circuit diagram of the IBSC containing seven defect species showing: (a) energy gaps and generated currents and (b) an equivalent circuit in terms of their associated chemical potential

The detailed balance theory assumes unity quantum efficiency in the limit so that the current coming out of a device is due to the difference between the incoming and outgoing photon flux. The current density due to the difference between the incoming and outgoing photon flux emitted by a blackbody surface per unit surface area over the energy range $E_l$–$E_h$ is:

$$J = q[f_s \dot{N}_s(E_l, E_h, T_s, 0) - f_c \dot{N}_c(E_l, E_h, T_c, u)] \quad (13)$$

According to the common Planck formula, the number of photons with energy between $E_l$ and $E_h$ which are emitted to the hemisphere at temperature $T$ by the blackbody is:

$$\dot{N}(E_l, E_h, T, u) = \frac{2An^2\pi}{h^3c^2} \int_{E_l}^{E_h} \frac{E^2 dE}{e^{(E-u)/kT}-1} \quad (14)$$

Here $q$ is the electronic charge; $h$ is Planck's constant; $c$ is the speed of light; $k$ is Boltzmann's constant; $E_l$ is the lower energy limit for the absorption process of interest; $E_h$ is the upper energy limit; $T_s$ is the temperature of the sun (modeled as a blackbody at a temperature of 6000K); $T_c$ is the temperature of the cell (modeled at a temperature of 300K) and $u$ is the chemical potential associated with the emitted radiation [18].

For a given external voltage $V$ to the solar cell, there is an equation:

$$qV = u_{CI} + u_{IV} = u_{CV} \tag{15}$$

It is now possible to write down the equations describing the current due to carriers being excited from the valence band to the *k*th defect [Eq. (16)], the current due to the carriers being excited from the *k*th defect to the conduction band [Eq. (17)] and the current due to the carriers being excited from the valence band to the conduction band [Eq. (18)].

$$J_{Vk} = q[f_s \dot{N}_s(E_k, E_{k-1}, T_s, 0) - f_c \dot{N}_c(E_k, E_{k-1}, T_c, u_{Ck})] \tag{16}$$

$$J_{kC} = q[f_s \dot{N}_s(Eg-E_k, Eg-E_{k+1}, T_s, 0) - f_c \dot{N}_c(Eg-E_k, Eg-E_{k+1}, T_c, u_{kC})] \tag{17}$$

$$J_{VC} = q[f_s \dot{N}_s(Eg, +\infty, T_s, 0) - f_c \dot{N}_c(Eg, +\infty, T_c, u_{VC})] \tag{18}$$

According to the assumption (C): $J_{Vk}=J_{kC}$, and with the Eq. (15), we can work out the chemical potential $\mu_{ck}$ or $\mu_{kV}$. Each absorbed photon only excites a single electron. The output power of the device is calculated as follows:

$$P_{out} = u_{vc}(J_{VC} + J_{D1} + J_{D2} + J_{D3} + J_{D4} + J_{D5} + J_{D6} + J_A) \tag{19}$$

The efficiency $\eta$ is defined as the ratio of the electrical power out over the solar power in. If the sun is modeled as a blackbody at a temperature of $T_s$, then the incident power received by the device is given by $p_{in} = f_s \sigma T_s^4$, $\sigma$ is the Stefan-Boltzmann constant.

Then the efficiency for solar cells is: $\eta = P_{out}/P_{in}$

The incident spectrum which we use is AM1.5 radiation where $f_s=f_c=1$. And the valence band and conduction band are unrestricted.

Calculating by the formulas above we get the relationship between solar cell conversion efficiency and the external voltage when different oxygen and sulfur impurity levels are introduced into Silicon [Fig. 6]. We can see that the maximum theoretical conversion efficiency of Silicon solar cell after introducing seven impurity levels is 51%, while the maximum theoretical conversion efficiency of common Silicon with unrestricted valence band and conduction band is 40.5%. The theoretical conversion efficiency of Silicon solar cell after introducing seven impurity levels is just 10% higher than that of common unrestricted Silicon under different external voltage. Some studies of our group indicate that the broad spectrum absorption efficiency of microstructured Silicon solar cell is very high [19, 20]. Obviously, the photoelectric conversion efficiency of femtosecond laser microstructured Silicon is not as high as its broad spectrum absorption efficiency, so microstructured Silicon cannot meet the requirements of high efficient IBSC according to the present findings. Therefore, the effect of defects in femtosecond laser microstructured Silicon on its properties must be further studied to reduce the negative impact from defects. Some relevant researches to minimize the negative impact of these defects in microstructured Silicon has been carried out [21].

Researches show that there are amounts of oxygen in Silicon. Its amount is determined by crystal growth and its existing form and content changes with annealing temperature and time. Its existence in ordinary Silicon is also inevitable, and it can form a donor level and an acceptor level. Now, we do not consider the impurity levels introduced by oxygen. According to the above formulas which only contain sulfur impurity levels, we get the relationship between theoretical conversion efficiency of solar cells and the external voltage (Fig.6). The Fig.6 shows that the conversion efficiency of IBSC increases by 10% or so based on the above results. Thus, microstructured Silicon after some certain machining process can make into high efficiency IBSC.

As the understanding for femtosecond laser microstructured Silicon is limited and the mechanism of its broad spectrum absorption is still under exploration, we have no idea whether this high conversion efficiency can be achieved by femtosecond laser-doped Silicon. It is known that impurities with concentration far higher than its solid solubility are introduced into Silicon by femtosecond laser irradiation and such a high concentration of impurities most likely formed continuous rather than discrete energy levels in Silicon. This could be proved by the fact that there is no peak in the absorption characteristic curve for microstructured Silicon. In addition, the high doping effect caused by high concentrations of impurity doping can not be ignored because it would severely affect the solar cell efficiency. High concentrations of doping would lead to shrinking of forbidden gap, and the shrinkage will decrease voltage of solar cells, thereby reducing efficiency; high doping effect would also lead to distinct decrease of carrier mobility and thus shorter minority carrier lifetime. However, for the nanostructure heavily doped by chalcogenides and deep-level sub-bands have been formed, it is possible for Auger effect to develop in the positive direction, that is, the electronic excitation probability is greater than the recombination probability, making a high-energy photon in microstructured Silicon be able to stimulate more than one electron, thus quantum efficiency higher than 100% could be obtained.

Besides the formation of stable impurity levels to restrain the nonradioactive recombination, for femtosecond laser microstructured Silicon, not only a large number of impurities but also abundant defects are introduced into Silicon. These defects, with impurities, will form amounts of recombination centers within the forbidden band, and these centers will increase the bulk recombination and severely effect the lifetime of minority carrier; There must be substantive defects and dangling bonds on the conical surface of femtosecond laser microstructured Silicon, which would speed up the surface recombination carrier; The high density of majority carriers in the surface layer may increase the Auger electron recombination caused by lattice collisions and shorten the lifetime of carrier. Thus the loss of carrier recombination most likely leads to electrical losses in solar cells based on femtosecond laser microstructured Silicon. So we expect to improve laser doping or other doping techniques to effectively control the impurity band position and width, and reduce the introduced defects in the doping process in the future, thus making maximized use of the strong wild spectrum absorption of black Silicon and making high efficiency IBSC with femtosecond laser microstructured Silicon.

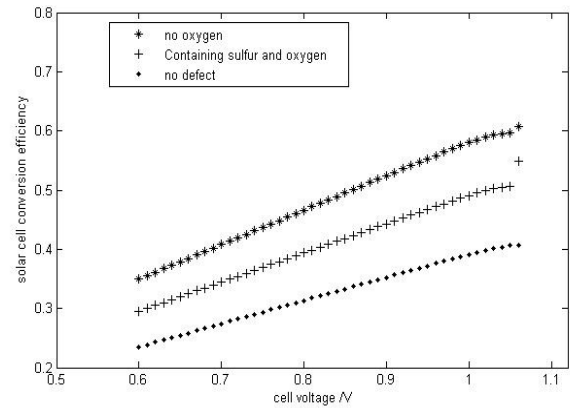

FIG.6. Relationship of solar cell conversion efficiency and the external voltage when doping sulfur with and without oxygen together into Silicon, Compared with no impurity level in Silicon forbidden band

IV. CONCLUSIONS

Solar cell with multi intermediate bands provides a completely new idea for making high efficient solar cells. and the concept of IBSC also arises at the historic moment. Although the study of IBSC is mostly confined to theoretical calculations and a small number of doping tests at present, however, IBSC attracts more and more attentions of scientific and technical workers because of its obvious advantages of cost and production. The further study of deep level impurity based on the mechanism of photoelectric conversion and the continuous improvement of technology in making impurity levels will promote a rapid development of the IBSC. Therefore, thorough study on the effect of high doping to femtosecond laser microstructured Silicon and the control of the position of impurity level are very important in making high efficiency solar cells. In this paper, based on the statistically relationship of the two neighboring thermal equilibrium occupancy and charge neutrality condition, we derive the nonlinear equation of the Fermi energy (this formula is universal) under the model of impurity level of microstructured Silicon when both donor and acceptor level exist at the same time. And the effect of Temperature and the Density of Impurities on Fermi level in the microstructured Silicon are shown in Fig.3-4. Then the theoretical conversion efficiency of the corresponding IBSC under our model is discussed by using the Detailed Balance Theory. We have discussed the effects of the impurity levels introduced by both oxygen and sulfur and only introduced by sulfur in microstructured Silicon on the conversion efficiency of IBSC. From Fig.6, we can see that the conversion efficiency of IBSC without the impurity levels introduced by oxygen increased by 10% or so than that with the impurity levels introduced by oxygen. This shows that the impurity levels introduced by oxygen in Silicon have a negative impact on the conversion efficiency of IBSC. So selecting the appropriate doping impurities and making clear the effect of the impurity and its position in the material on solar cell efficiency are absolutely necessary for high efficiency solar cells.